\documentstyle[11pt]{article}
\newcommand{\blankline}{\vskip .3cm}
\newcommand{\f}{\begin{equation}}
\newcommand{\ff}{\end{equation}}

\begin{document}
\centerline{\LARGE Covariant quantization}
\blankline
\rm
\centerline{\LARGE of membrane dynamics}
\centerline{Lee Smolin${}^*$}
\blankline
\centerline{\it  Center for Gravitational Physics and Geometry}
\centerline{\it Department of Physics}
 \centerline {\it The Pennsylvania State University}
\centerline{\it University Park, PA, USA 16802}
 \blankline
 \blankline
 \blankline
\centerline{October 15, 1997}
 \blankline
 \blankline
 \blankline
\centerline{ABSTRACT}
A Lorentz covariant quantization of membrane dynamics is 
defined, which also leaves unbroken the full three dimensional
diffeomorphism invariance of the membrane.  This makes it possible
to understand the reductions to string theory directly in terms of the
Poisson brackets and constraints of the theories.   Two approaches
to the covariant quantization are studied, Dirac quantization and
a quantization based on matrices, which play a role in recent work
on $\cal M$ theory.  In both approaches the dynamics is generated
by a Hamiltonian constraint, which means that all physical states
are ``zero energy".    A covariant matrix formulation may be defined,
but it is not known if the full diffeomorphism invariance 
of the membrane may be consistently  imposed.  The problem
is the non-area-preserving diffeomorphisms: they are realized
non-linearly in the classical theory, but in the quantum theory they
do not seem to have a consistent implementation for finite N.  
Finally, an approach to a genuinely background independent 
formulation of matrix dynamics is briefly described.

\blankline
${}^*$ smolin@phys.psu.edu
\eject

\section{Introduction}

This paper studies the quantization of a theory of the embeddings
of membranes in $d$ dimensional spacetime, using methods that
preserve the manifest Lorentz invariance of the theory.   This
problem is of interest first of all because the quantum theory
of the super-membrane in $10+1$ dimensions\cite{smem} 
is intimately 
associated with current attempts to construct $\cal M$ theory,
which is a conjectured non-perturbative formulation of string
theory\cite{mtheory}.  The quantum theory of the 
supermembrane in the light cone gauge is known\cite{DHN}.
This gauge fixed 
version of the theory is elegantly described in terms of
a theory of $N \times N$ matrices.  One of the issues the present
paper attempts to answer is whether there is a covariant
quantization of the membrane that is also expressed in terms
of matrices.

This is an important question because the same matrix quantum
theory has been conjectured to give a description of $\cal M$ theory
in the infinite momentum frame\cite{BFSS}.    
This conjecture is motivated
by another intriguing fact, which is that the same matrix quantum
theory can be obtained as the dimensional reduction of 
supersymmetric quantum mechanics to zero spatial dimensions.

It is of great interest to know to what extent this triple
correspondence, between the supermembrane, 
the reduction of supersymmetric
quantum mechanics and $\cal M$ theory is restricted only
to the light cone gauge and the infinite momentum frame, or
has a larger range of validity.  To answer this question we
need formulations of these theories which are
Lorentz covariant.    

In the present paper we study these problems by making a
canonical quantization of membrane dynamics which is Lorentz
covariant in the background spacetime. This turns out to be
straightforward, so long as no gauge conditions are fixed on the
membrane itself.  This leads to a canonical formulation of the
membrane dynamics that has both manifest Lorentz covariance
and complete invariance under the diffeomorphism group of the
membrane\footnote{This formalism has been sketched, but
less completely, in \cite{early}.}.   For completeness we include
also the coupling to the three form field $A_{\alpha \beta \gamma}$.

Beginning with this classical 
formalism, which is
set out in the next section, we then study two different approaches
to the quantization: Dirac quantization and quantization in terms of
matrices as in the light  cone gauge fixed theory.  

One limitation of the present study is that 
most of the results reported below 
hold for any dimension $d$, and 
we have not so far completed the extension to the supermembrane.
The extension to the $10+1$ dimensional supermembrane is
expected to be straightforward, and will be carried out elsewhere.
While we expect further insight from this extension, several of the  
results already found do apply directly to the $10+1$ dimensional
supermembrane and are of immediate relevance
to the question of
the Lorentz covariant form of $\cal M$ theory.  Among them are: 

\begin{itemize}

\item{}The dynamics 
is given in terms of a Hamiltonian constraint, which is similar
to the Hamiltonian of the light-cone gauge fixed formalism, 
except that all $d$ matrices are present and the Lorentz metric
ties up the indices, as a result of the manifest Lorentz covariance.

\item{}There are two important consequences of the fact that the dynamics
is given by a constraint.  The first is that all states 
are zero energy and the second is that as a consequence physical
states are expected to be non-renormalizable in the naive inner
product.  A new inner product on physical states must be chosen.
This has implications for the construction and interpretation of
zero energy states\footnote{These are discussed in the
light-cone gauge fixed theory in \cite{zero}.}.

\item{}A Lorentz covariant quantization in terms of matrices is
given at a formal level (which means that the limit 
$N\rightarrow \infty$ is not understood) in section 6.  
There are two 
more matrices in the Lorentz covariant formulation than in the
light cone gauge fixed formalism, but these are 
balanced by two new sets of constraints.  One is the Hamiltonian
constraint, the other is a set of the two dimensional diffeomorphism
group of the constant time surfaces of the membrane that is broken
in the light-cone gauge fixed formalism.  These are the 
area-non-preserving diffeomorphisms.  If the theory is going to
be formulated in terms of matrices these must be realized
non-linearly.  I show that this can be done at the classical level,
which tells us how to do it formally in the quantum theory.  But
whether it can actually be done depends on issues of 
regularization and operator ordering that have not yet been
resolved.   It is likely that these gauge symmetries can only
be consistently imposed in the $N \rightarrow \infty$
limit.  This is the main difficulty that must be solved if
there is to be a Lorentz covariant formulation in terms of
matrices.

\item{}It has been known for
sometime that there is a limit in which the membrane theory
reduces to the string theory\cite{limitstrings}.  
I show in section 3 that this can be 
understood completely at the level of the phase space and 
constraints of the canonical theory.

\item{}In the special case of $2+1$ dimensions we can find
an exact physical state of the theory, which 
is an analogue of the Chern-Simons state\cite{kodama}
that plays an important role in quantum 
gravity\cite{kodama,chopinme}.  I show in section
5 how this may be interpreted
as a semiclassical state associated to a certain class of solutions. 
This provides further evidence for the physical character of
zero-energy states that are non-normalizable under the naive
inner product\footnote{A similar state is studied in the 
$7$ dimensional theory in \cite{DHN}.}.

\end{itemize} 

Before closing the introduction we should remark that
no Lorentz covariant theory can be more than
a step on the road to the true, non-perturbative form of 
$\cal M$ theory.  Whatever it is, we know that $\cal M$ theory 
cannot have its
most fundamental formulation in terms of fields, strings, membranes
or anything else moving in a classical spacetime manifold.  This is
so because $\cal M$ theory must be a non-perturbative theory of
quantum gravity and in any such theory the geometry of spacetime 
must emerge from a more fundamental quantum system that is not
dependent on any background spacetime for its description.  Such
a theory may have gauge invariances such as diffeomorphism invariance
or some extension of it; what it cannot have is any global symmetries
that depend on the geometry of fixed background metrics.

Thus, the key question in $\cal M$ theory is to find its 
background-independent, non-perturbative formulation.  
In section 7 a few steps towards such
a theory is taken.  I show that there are lagrangian and
hamiltonian formulations of matrix dynamics in which the
global symmetries are replaced by a matrix valued extension
of diffeomorphism invariance.  The relationship of this theory
to $\cal M$ theory is, however, unknown.

\section{Hamiltonian reduction without gauge fixing}

We begin with the action for a $2+1$ dimensional membrane
$\cal M$
embedded in $d$ dimensional Minkowski spacetime in interaction
with a three form field $A_{\alpha \beta \gamma}$.
\f
S= \int_{\cal M} \sqrt{-g}  + e \int_{\cal M} A
\ff

Here $g=(detg_{ij})$ where $g_{ij}$ is the induced metric
given in terms of the embedding coordinates
$X^\alpha (t , \sigma^A )$ by
\f
g_{ij}= \partial_i X^\alpha \partial_j X^\beta \eta_{\alpha \beta}
\ff

Here $\eta_{\alpha \beta}$ is the Minkowski metric of the
$d$ dimensional background spacetime, so that 
$\alpha , \beta = 0,...,d-1$ and the three coordinates of the
worldsheet, $\sigma^i$, $i=0,1,2$ are broken down into 
$\sigma^0=\tau $ and $\sigma^A$, $A=1,2$.   $\eta^{\alpha \beta}$
will be used to raise and lower spacetime indices. 

The action can then be written
as
\f
S=\int_{\cal M} 
\sqrt{{1\over 2} \dot{X}^\alpha \dot{X}^\beta G_{\alpha \beta}(X)}
 + e \int_{\cal M} A
\label{act1}
\ff
where $G_{\alpha \beta}$ is  a metric on the 
configuration space of the embeddings $X^\alpha (\sigma^A)$,
which is given by
\f
G_{\alpha \beta}= \eta_{\alpha \beta}h -2h_{\alpha \beta}
\ff
where $h=\eta^{\alpha \beta}h_{\alpha \beta}$ and 
$h_{\alpha \beta}$ is the useful quantity
\f
h_{\alpha \beta}= < X_\alpha ,X_\gamma ><X_\beta , X^\gamma >
\ff
where $< X_\alpha ,X_\gamma >$ denotes the ``manifold Poisson
bracket",
\f
< X_\alpha ,X_\gamma >=\epsilon^{AB}
\partial_A X_\alpha \partial_B X_\gamma
\ff
This of course has nothing to do with the phase space Poisson
bracket we will shortly introduce.  

We may note that the action is in the Barbour-Bertotti form \cite{BB}
which shows that there is no intrinsic preferred time variable on
the membrane.  It also gives us an interpretation of the theory.
We assume the topology of the membrane is fixed 
to be $\Sigma \times R$,
with $\Sigma$ a compact two manifold. Then the 
configuration space $\cal C$ 
of the membrane consists of the embeddings of  
the two manifold $\Sigma$ into $d$ dimensional
Minkowski spacetime.  In coordinates this is given by 
$X^\alpha (\sigma )$.  $\cal C$ is an infinite dimensional
manifold that has on it an indefinite metric given by
$G(\delta_1 X^\alpha , \delta_2 X^\beta )
= \int_\Sigma G_{\alpha \beta}\delta_1 
X^\alpha \delta_2 X^\beta$.
Then the action (\ref{act1}) tells us that 
when $A_{\alpha \beta \gamma}=0$ the histories of the
membrane trace out timelike geodesics of $G$.

This Barbour-Bertotti form also tells us how to construct the unconstrained
Hamiltonian formulation, following the procedure used
for that theory\cite{BB-me}.  
Without doing any gauge fixing we proceed directly
to find the canonical momenta\footnote{We use signature
$+----...$},
\f
p_\alpha (\sigma ) = {\partial S \over \partial 
\dot{X}_\alpha (\sigma )} = {1\over \sqrt{-g}} G_{\alpha \beta}
\dot{X}^\beta 
+ e A_{\alpha \beta \gamma}<X^\beta , X^\gamma >
\label{pdef}
\ff
The elementary Poisson brackets are,
\f
\{ X^\alpha (\tau, \sigma^A ) , p_\beta (\tau , \sigma^{A\prime} )
\} = \delta^2 (\sigma , \sigma^\prime ) \delta^\alpha_\beta
\label{pb}
\ff

We find immediately three primary constraints, which
follow only from the  definition of the momenta
(\ref{pdef}).
The diffeomorphism constraints are
\f
D_A (\sigma ) = (\partial_A x^\alpha ) p_\alpha  -
e A_{\alpha \gamma \delta }<X^\gamma , X^\delta > =0
\label{diffeo}
\ff
and it is easy to see that acting on the embedding coordinates
$X^\alpha (\sigma )$ they generate $Diff(\Sigma )$.  Then
there is the Hamiltonian constraint
\f
{\cal H}(\sigma ) ={1 \over 2}  G^{-1 \alpha \beta}
(p_\alpha -e A_{\alpha \gamma \delta }<X^\gamma , X^\delta > )
(p_\beta -e A_{\beta \rho \sigma }<X^\rho , X^\sigma > )
-1 =0
\label{ham1}
\ff
where $G^{-1 \alpha \beta}$ is the inverse of $G_{\alpha \beta}$,
i.e. $G^{-1 \alpha \beta} G_{\beta \gamma}= \delta^\alpha_\gamma$.
It is straightforward to verify that the vanishing of these constraints
follows from the definition of $p_\alpha$.

$G^{-1 \alpha \beta}$ may be constructed as a power series as
\f
G^{-1 \alpha \beta}= {1 \over h} \left (\eta^{\alpha \beta} + 
{2 h^{\alpha \beta} \over h} +...  \right ) 
 \ff
However, the nonlinear terms actually do not affect the
evolution on the constraint surface because
\f
h^{\alpha\beta} p_\beta = 2det(q)q^{AC} (\partial_A x^\alpha )
D_C  .
\ff
Here $q_{AB}$ is the two dimensional induced metric
on $\Sigma$
defined by  $q_{AB} = \partial_A X_\alpha \partial_B X^\alpha$.
(Note that $h= {1\over 2} det(q)$ is negative as the induced
metric has Minkowskian signature.)
This means that
\f
{\cal H} =  {\cal H}_0 + D_A R^{AB} D_B
\label{oh}
\ff
where we have thus a new linear combination of constraints
\f
{\cal H}_0 = {1\over 2} p_\alpha
{\eta^{\alpha \beta} \over h} p_\beta
-1
\label{oh2}
\ff
and
\f
R^{AB}= {2 \over h} qq^{AB} +...
\label{oh3}
\ff
Thus, on the diffeomorphism constraint surface $D_A=0$
we have
\f
{\cal H} \approx {\cal H}_0
\ff

It is easy to verify that these constraints close to
give an algebra very like the 
$2+1$ dimensional $ADM$ algebra,
\f
\{ D(v) ,D(w) \} = D([v,w])
\label{diffalg}
\ff
with $D(v) =\int_\Sigma v^A D_A$.  We also have
\f
\{ {\cal H}_0(N) , {\cal H}_0 (M) \} =
\int_\Sigma \left (
M\partial_A N- N\partial_A M  \right )
{{\cal H}_0 \over h} qq^{AB} D_B
\ff
and
\f
\{ D(v) , {\cal H}_0(N) \} = {\cal H}_0 ({\cal L}_v(N))
\label{dh}
\ff

Finally, it is convenient to densitize the constraint (\ref{oh2})
to make the constraint polynomial, which gives us\footnote{
We employ, inconsistently, the convention that density weights
are marked with a tilde.  Note that the restriction of
$\tilde{\tilde{\cal H}}_0$ to the transverse coordinates is
minus the usual light cone gauge Hamiltonian.},
\f
\tilde{\tilde{\cal H}}_0 = h {\cal H}_0 = 
{1\over 2} p_\alpha
\eta^{\alpha \beta}  p_\beta -h
\label{nod}
\ff

\section{Relation of the membrane to string theory at the
classical level}

It is easy to demonstrate within the canonical
framework that string theory may be recovered
from a particular limit of membrane theory.  Let us consider
a membrane whose spatial sections have topology
$S^1 \times S^1$ with coordinates $\sigma$ on the first $S^1$ and
a periodic coordinate $\rho \in [-1,1]$ on the second $S^1$, which
satisfies $\oint d\rho = 1$.  We than take 
an ansatz for an evolution of a membrane in $D$ dimensional
spacetime of the form,
\f
X^\alpha (\tau, \sigma, \rho ) = Z^\alpha (\tau , \sigma )
+ \epsilon \rho  W^\alpha (\tau , \sigma )
\label{ansatz}
\ff  
What we are doing is reducing the embeddings of the membrane
$\cal M$ to the embedding of a worldsheet $\cal S$ defined
by the condition $\epsilon=0$.  
$Z^\alpha$ is the embedding coordinate
of the worldsheet and it and $W^\alpha$ are then fields on the
worldsheet.  In the limit $\epsilon \rightarrow 0$ the 
membrane goes over into the worldsheet; we want to
see if the dynamics of the membrane goes over into
bosonic string theory in the same limit.  To accomplish this
we need a condition on the $W^\alpha$ so that they
are restricted by the $Z^\alpha$.  To see what it should
be we  look at
the action for the membrane in the presence of the 
ansatz (\ref{ansatz}) and in limit of
small $\epsilon$.  We have, for $I,J,K=\tau, \sigma$ coordinates
on the worldsheet.  
\f
det(g_{ij})= \epsilon^2  \left (
det(q_{IJ}) W_\alpha W^\alpha -2det(q_{IJ}) q^{KL} (W_\alpha 
\partial_K Z^\alpha )(W_\beta 
\partial_L Z^\beta )
\right ) +O(\epsilon^3 )
\ff
where $q_{IJ}$ is the induced metric on $\cal S$.  
(This should be distinguished from $q_{AB}$, the induced
metric on the constant time surfaces of the membrane,
which we called $\Sigma$.)
Hence we see that
the conditions we require are 
\f
W_\alpha W^\alpha=1 , W_\alpha \partial_\sigma Z^\alpha= 
W_\alpha \partial_\tau Z^\alpha= 0
\label{ansatz2}
\ff
so that
\f
 det(g_{ij})= \epsilon^2 det(q_{IJ})  + O(\epsilon^3 )
\ff
Given the $Z^\alpha (\tau, \sigma )$ this gives three equations
at each point $(\tau, \sigma )$  
to determine the three $W^\alpha (\tau, \sigma )$.
Once we have done this we have that
\f
S^{membrane}= \epsilon  \int d\tau d\sigma \sqrt{-det(q_{IJ}) }
= \epsilon S^{Nambu}(Z^\alpha )
\ff
This correspondence goes through in the equations of
motion as well.  To show this we look at the definition of the
momenta for the membrane (\ref{pdef}), which gives us
\f
p_\alpha = \epsilon 
{(\partial_\sigma Z^\alpha )^2 \over \sqrt{det(q)}}
\eta_{\alpha \beta}\dot{Z}^\beta + O(\epsilon )
\ff
On the other hand, the definition of the momentum of
the string, (in the presence of the momentum constraint,
$\partial_\tau Z^\alpha \partial_\sigma Z_\alpha =0$) is
\f
p^{str}_\alpha = 
{(\partial_\sigma Z^\alpha )^2 \over \sqrt{det(q)}}
\eta_{\alpha \beta}\dot{Z}^\beta
\ff
Thus, we have
\f
p^{str}_\alpha (\tau , \sigma ) = \lim_{\epsilon \rightarrow 0}
{1 \over \epsilon }  \oint_{S^1} d\rho 
p_\alpha (\tau, \sigma ,\rho ) 
\label{crucial}
\ff
From the densitized hamiltonian constraint, (\ref{nod}), we 
have for the potential energy of the membrane
\f
{\cal V}^{mem}= < X_\alpha , X_\beta ><X^\alpha , X^\beta >
= {2 \epsilon^2 \over l_{pl}^2} (\partial_\sigma Z^\alpha )^2 +
O(\epsilon^{3})
\ff
Thus, the Hamiltonian constraint of the membrane theory is
of the form
\f
\tilde{\tilde{\cal H}}_0= \epsilon^2 {\cal H}^{string} +
O(\epsilon^3 )
\ff
where, 
\f
{\cal H}^{string}= {1\over 2}p^{str}_\alpha p^{str}_\beta
\eta^{\alpha \beta} + (\partial_\sigma Z^\alpha ) 
(\partial_\sigma Z^\alpha ) \eta_{\alpha \beta}
\ff
Similarly, we have for the diffeomorphism constraints,
\f
D(v)= \int d\sigma \oint_{S^1 } 
d\rho v^A p_\alpha \partial_A X^\alpha (\sigma ,\rho ) =
\epsilon \int d\sigma  v^\sigma p_\alpha^{str} \partial_\sigma Z^\alpha (\sigma ) = \epsilon D(v^\sigma)^{string}
\ff

Thus, we have shown that there is a limit in which a sector
of the membrane theory goes over into the bosonic theory.

\section{Realization of the two dimensional diffeomorphisms}

One goal of the present work is to construct a quantization of the
Lorentz covariant Hamiltonian dynamics described here in
terms of a matrix representation similar to that used in the 
light-cone gauge
fixed formalism.  The main obstacle to doing this is that it is
only the subgroup of the two dimensional diffeomorphism group
that preserve the area element of the induced metric that
are represented in the matrix formalism by $SU(N)$ transformations,
in the limit of large $N$.  This is fine for the light-cone gauge fixed
formalism, because there the full diffeomorphism group of 
$\Sigma$ has been broken down to the area 
preserving ones\cite{DHN}.
But if we want to quantize the covariant formalism we have to
represent all of $Diff(\Sigma )$.  In order to understand how to
do this we must first study how the non-area preserving 
diffeomorphisms act on the embedding coordinates and momenta.

To do this we split
the vector fields $v^A$ into the area preserving and 
area-non-preserving part, each of which is given by a 
scalar field. We call them $a$ and $n$ for area preserving and
non-area preserving.  The decomposition is
\f
v^A = {1 \over \sqrt{q}}\epsilon^{AB}\partial_B a
+ q^{AB}\partial_B n
\ff
We have
\f
{\cal L}_v \sqrt{q}= \partial_A \sqrt{q} v^A = \sqrt{q} 
\nabla^2 n
\ff
where
\f
\nabla^2 = {1 \over \sqrt{q}} \partial_A \sqrt{q}q^{AB} \partial_B
\ff
showing that $a$ parameterizes the area preserving subgroup
of $Diff(\Sigma )$, which we call $Diff_{\sqrt{q}}(\Sigma )$
while $n$ parameterizes the coset 
$Diff(\Sigma )/Diff_{\sqrt{q}}(\Sigma )$.  

The action of the area preserving part defines a vector
density $\tilde{a}^A= \epsilon^{AB}\partial_Ba$
whose action on functions is embedded in the Poisson
algebra of functions
\f
\tilde{a}^A\partial_A \phi = <\phi , a >
\label{lineara}
\ff
Thus, the map $\phi : \tilde{a}^A \rightarrow a$ of divergence
free vector fields to scalars defines an embedding of the
Lie algebra of area preserving diffeomorphisms into the
Poisson algebra on $\Sigma$ given by $<,>$.  It is this
Poisson algebra which is mapped to $SU(N)$ in the limit
$N \rightarrow \infty$ in the quantization of the membrane
in which the embedding coordinates $X^\alpha (\tau , \sigma ,\rho )$
are mapped to matrices $X^{\alpha I}_{\ \ J}$ \cite{DHN}.

What about the non-area preserving part?  This is given also by
functions, but the action does not map linearly into the Poisson
algebra on $\Sigma$.  However, we can find a non-linear
realization of the generators of 
$Diff(\Sigma )/Diff_{\sqrt{q}}(\Sigma )$ on the embedding
coordinates $X^\alpha$ and their conjugate momenta
$\tilde{p}_\alpha$.    If we consider the undensitized non-area
preserving vector field,
\f
N^A=q^{AB}\partial_B n
\ff
then using the definition of the induced metric 
we have for any function $f$ and density $\tilde{\omega}$ on
$\Sigma$
\f
{\cal L}_N f = { < f, X^\alpha ><n , X_\alpha > \over
< X_\mu , X_\nu ><X^\mu , X^\nu > }
\ff
\f
{\cal L}_N \tilde{\omega}= \partial_A ( \tilde{\omega} N^A ) =
<  \tilde{\omega} { <n , X_\alpha > \over
< X_\mu , X_\nu ><X^\mu , X^\nu > } ,  X^\alpha >
\ff
These equations apply, in particular to the $X^\alpha$
and $\tilde{p}_\alpha$ (which is, of course, a density on
$\Sigma$.)  The first gives a non-linear
realization of $Diff_q (\Sigma ) / Diff(\Sigma )$.
\f
{\cal L}_N X^\beta = { < X^\beta , X^\alpha ><n , X_\alpha > \over
< X_\mu , X_\nu ><X^\mu , X^\nu > } .
\label{xlaw}
\ff
The second gives the transformation of the momenta
\f
{\cal L}_N \tilde{p}_\alpha = \partial_A ( \tilde{p}_\alpha N^A ) =
<  \tilde{p}_\alpha { <n , X_\beta> \over
< X_\mu , X_\nu ><X^\mu , X^\nu > } ,  X^\beta >
\label{plaw}
\ff
As these are diffeomorphisms, by (\ref{dh}) they must
leave the constraint surface ${\cal H}_0=0$ invariant.
Thus, the theory has {\it two} gauge invariances, each given
by a mapping of $Diff(\Sigma ) $ into the algebra of functions
on $\Sigma$.  The first is the linear action (\ref{lineara}) of
the area preserving transformations.  The second is the
non-linear representation of 
$Diff(\Sigma )/Diff_{\sqrt{q}}(\Sigma )$ which is given by
(\ref{xlaw}) and (\ref{plaw}).  Both must be represented
in a quantization of the covariant theory.

\section{Dirac Quantization}

We can now discuss the quantization of the membrane theory.  
I will discuss briefly two methods
of quantization.   We start with Dirac quantization.  This
is straightforward, but makes so far no connection with the
matrix models.  We do find one interesting result which is
that in
the particular case of $2+1$ dimensions  we can find an
exact physical state that describes the  reduction of the
membrane to the string.  After describing this we will
turn to the question of the existence of a matrix 
representation of the covariant membrane.

Under the procedure of Dirac quantization one begins with some
kinematical hilbert space ${\cal H}^{kin}$ and establishes
the canonical commutation relations associated to the
Poisson brackets (\ref{pb}).  The natural representation to use is
the configuration space representation $\Psi [X^\alpha ]$,
where the kinematical configuration space ${\cal C}^{kin}$ consists
of maps $X^\alpha (\sigma , \rho ) : \Sigma \rightarrow M^N$ 
from the two surface $\Sigma$ to $N$ dimensional Minkowski
spacetime.  The operator assignments are the natural ones in
which
\f
p_\alpha \Psi = \imath \hbar {\delta \Psi \over \delta X^\alpha}
\ff
On this we impose first the diffeomorphism constraints (\ref{diffeo})
in the form
\f
\hat{D}(v)\Psi [X^\alpha ] = \int_\Sigma ({\cal L}_V X^\beta ) 
{\delta \Psi \over X^\beta }[X^\alpha ]
\ff
This is solved in general by the requirement that 
\f
\Psi [X^\alpha ] = \Psi [ \phi \circ X^\alpha ]
\ff
where $\phi \in Diff(\Sigma )$ so that the states become
functionals on ${\cal C}^{diffeo} ={\cal C}^{kin}/Diff(\Sigma )$.
The problem is then to invent a regularization so that 
the solutions to 
\f
{\cal H} (N) \Psi =0
\ff
can be extracted.  Once this is done a physical inner product is
to be picked on the space of solutions to both sets of constraints.

In particular cases some exact solutions can be found.  For example,
for the case of $N=3$ we can split the Hamiltonian into self-dual
and anti-self-dual parts
\f
H_0 = {1 \over 2} P^-_\alpha P^{+ \alpha}
\ff
where
\f
P_\alpha^\pm = p_\alpha \pm  
\epsilon_{\alpha \beta \gamma } <X^\beta , X^\gamma > 
\ff
An analogue of the Chern-Simons state for quantum gravity
\cite{kodama} can be construct using
\f
Y[X^\alpha ] = {1\over 3} \int_\Sigma 
\epsilon_{\alpha \beta \gamma } X^\alpha <X^\beta , X^\gamma > 
\ff
so that,
\f
{\delta Y \over \delta X^\alpha } = 
\epsilon_{\alpha \beta \gamma } <X^\beta , X^\gamma > 
\ff
If we define the ``Chern-Simons state"  by
\f
\Psi_{CS}[X^\alpha ] = e^{\imath Y[X^\alpha ]}
\label{cs}
\ff
it follows directly that
\f
P^+_\alpha \Psi_{CS} [X^\alpha]=0
\ff
Since this state is manifestly invariant under
$Diff(\Sigma )$ this is a well defined physical state.

It may be objected that the state is not-normalizable.  However,
this is only the case in a naive Fock inner product, which might
be established on the kinematical state space ${\cal H}^{kin}$.  This
objection rules out the consideration of an analogous state in the
case of Yang-Mills theory.
However, this objection does not hold in the case of theories
whose dynamics is governed by constraints, because 
{\it all} physical
states, being zero energy states of the Hamiltonian {\it constraint}
are expected to be non-normalizable in this kinematical inner
product.  The inner product on physical states must be
constructed on the space of solutions to the constraints.
Since we do not have a full space of physical states we are not
yet in a position to do this, on the other hand, at the present
stage there can be 
no objection to taking the state $\Psi_{CS}[X^\alpha ]$ to be
physical as a working hypothesis and seeing where it leads.  We
may note that in the case of quantum gravity there are good 
arguments that the analogous state is in fact the full 
non-perturbative vacuum state for the theory in the presence
of a cosmological constant.  In this case both the exact Planck scale
description and semiclassical limit are understood.  For small
cosmological constant the state has a semiclassical
interpretation which describes fluctuations around De Sitter
spacetime\cite{kodama,chopinme}, while the exact description 
of the state is as
the Kauffman invariant of quantum spin networks at level
$k=6\pi /G^2\Lambda$ \cite{mejmp}.

In fact the Chern-Simons state in the present context must
also have a semiclassical interpretation, since it is of the
form of a $WKB$ state.  To find that interpretation we note
that treating $Y[X^\alpha ]$ as a Hamilton Jacobi function we
have
\f
p_\alpha = {\partial Y \over \partial X^\alpha } = 
\epsilon_{\alpha \beta \gamma} < X^\beta , X^\gamma >
\ff
We may note that this satisfies the classical 
hamiltonian and momentum
constraints.  To find the velocities we may use the time defined
by the densitized hamiltonian constraint (\ref{nod}), so that
\f
\dot{X}_\alpha = \{ X_\alpha , \tilde{\tilde{\cal H}}_0 \}
= p_\alpha = \epsilon_{\alpha \beta \gamma} < X^\beta , X^\gamma >
\label{interesting}
\ff
The state (\ref{cs}) then is a semiclassical state that describes
fluctuations around the solutions to this equation. 

We may note that a similar state can be constructed in seven 
dimensions using the octonions \cite{DHN}, by replacing 
$\epsilon_{\alpha \beta \gamma}$ in (\ref{cs}) 
by the structure constants
for the seven imaginary octonions.  In fact, the
octonions can be used to give a compact expression to 
M(atrix) theory, which will be described in \cite{me8}.

\section{Is there a matrix formulation of the covariant theory?}

 It would be very convenient if the regularization of the
light cone gauge fixed theory in terms of $N \times N$ 
hermitian matrices
could be carried out as well for the covariant version of the
theory.  
To investigate this we may consider
states of the form $ \Psi[\hat{X}^\alpha ] $
where the $\hat{X}^\alpha $ are $d$ $N\times N$ hermitian
matrices in 
$d$ dimensional spacetime.  The momenta 
$\tilde{p}_\alpha$ are then represented as
$\partial / \partial \hat{X}^\alpha $.  The algebra of functions on
$\Sigma$ under $<,>$ is then taken over to the matrix algebra,
so that $ <X^\alpha , X^\beta > \rightarrow 
[\hat{X}^\alpha , \hat{X}^\beta ]$.  The area element
preserving subgroup of the 
diffeomorphism group $Diff_{\sqrt{q}}(\Sigma )$
then map to the group $SU(N)$, which becomes the gauge group.

This is sufficient for the light cone gauge theory, where the
area element preserving diffeos are the only gauge symmetry,
but will it work for the covariant formulation, where the 
gauge symmetry is expanded to the full
$3$ dimensional diffeomorphism group of the membrane?  To do
this we must implement on the $SU(N)$ invariant functionals
of the membrane two additional constraints, which are,
formally, the hamiltonian constraint,
\f
\hat{\cal H}_0 \Psi [\hat{X}^\mu ] = \left [
- {\partial^2  \over \partial \hat{X}^\alpha \partial \hat{X}_\alpha }
+ [\hat{X}^\alpha , \hat{X}^\beta ][\hat{X}_\alpha , \hat{X}_\beta ]
\right ] \Psi [\hat{X}^\mu ] =0
\label{ham9}
\ff
and the area non-preserving part of the diffeomorphisms
of $\Sigma$.    We may note that the counting is right;
this formalism has two more matrix degrees of freedom than 
the light cone gauged fixed theory, but these are balanced by
two additional matrix valued constraints.  
Presumably the Hamiltonian constraint can
be implemented, as it differs only by some signs
from the Hamiltonian operator that has been studied in the
light cone gauge fixed theory.  The difficulty is with the
remaining non-area preserving diffeomorphisms; at 
present the author is unaware of any method for
implementing them.

To have a chance of succeeding we can multiply the vector
field by $h$ to get  polynomial transformation laws.  
(This
step is implicit in writing the area preserving diffeomorphisms
in terms of $SU(N)$ transformations, so we use it here as well.)
Using symmetric ordering to preserve the hermiticity of the
matrices we find transformation laws of the form, 
\f
\delta \hat{X}^\mu = 
[\hat{n} , \hat{X}^\alpha ][\hat{X}_\alpha , \hat{X}_\mu ] + 
[\hat{X}_\alpha , \hat{X}_\mu ] [\hat{n} , \hat{X}^\alpha ]
\label{xlaw2}
\ff
\f
\delta \hat{p}_\alpha = 
[\hat{p}_\alpha [\hat{n} ,\hat{X}_\mu ] ,\hat{X}^\nu]
+ [ [\hat{n} ,\hat{X}_\mu ] \hat{p}_\alpha ,\hat{X}^\nu]
\label{plaw2}
\ff
Equivalently, up to an $SU(N)$ transformation these can be
replaced by a corresponding set of double commutator 
transformations,
\f
\delta \hat{X}^\mu =  
[[\hat{n},\hat{X}_\alpha ] , \hat{X}^\mu ] \hat{X}^\alpha +
\hat{X}^\alpha [[\hat{n},\hat{X}_\alpha ] , \hat{X}^\mu ] 
\ff
Acting on quantum states these should generate the constraint,
\f
\hat{\cal D}[\hat{n}] \Psi [\hat{X}^\rho ] = \left (
[[\hat{n},\hat{X}_\alpha ] , \hat{X}^\mu ] \hat{X}^\alpha +
\hat{X}^\alpha [[\hat{n},\hat{X}_\alpha ] , \hat{X}^\mu ] 
\right ) {\delta \Psi [\hat{X}^\rho ] \over \delta \hat{X}^\mu}
\ff
Unfortunately, at least for finite $N$,  
these do not appear to generate a symmetry of
the Hamiltonian constraint (\ref{ham9}).  
It  seems likely that if  these symmetries can 
be implemented exactly, it will be  only in the 
$N \rightarrow \infty$ limit\footnote{Djordje Minic has
kindly informed me that Hidetoshi Awata and he have considered
similar issues in the context of a covariant 
lagrangian matrix theory.}.   
It is also possible to speculate that this additional symmetry has
something to do with the ``hidden" symmetries in supergravity
and string theory, however there is little more that can be said
unless a way is found to implement them in the quantum theory.

\section{Towards a genuinely non-perturbative form of $\cal M$ 
theory}

Before closing this paper, we turn briefly to the key problem
of finding a fundamental, background
independent formulation of $\cal M$ theory.  Such
a formulation may have no dependence on a particular
classical spacetime.  Nor can it have any global symmetries, as
those arise in general relativity and other 
gravitational theories only as symmetries of
particular solutions.  A theory that has diffeomorphism invariance,
or some extension of it as the fundamental gauge symmetry
cannot have any global symmetries associated with particular
spacetime manifolds. 

This follows from general arguments about the role of
diffeomorphism invariance in theories in which the
spacetime geometry is a dynamical field.  Other arguments,
coming directly from string theory lead to the same 
conclusion.  For example,  $T$ duality and the other 
dualities tell us that string
theories defined as expansions around different spacetime
backgrounds are sometimes completely equivalent to each 
other\cite{duality}.
There are further more arguments that these dualities are to be
considered to be gauge symmetries of $\cal M$ theory. In that case
the gauge invariant description cannot be given in terms of
fixed classical backgrounds.

Whatever else it has accomplished,
the studies of non-perturbative quantum 
gravity\cite{lp1,volume1,future} and topological
quantum field theory, and their 
inter-relations\cite{future} have shown
us that it is possible to construct background independent,
diffeomorphism invariant quantum field theories,
even to the level of mathematical rigor reached by ordinary
constructive quantum field theory\cite{rigorous}.   
This should give us
the confidence to attempt the same for $\cal M$ theory.

One strategy to construct such a theory would be 
to construct a dynamics of $N \times N$ matrices which has
no global symmetries, but instead a group of gauge symmetries
larger than $SU(N)$.  
The simplest way to do this is to find an action which is a
functional of a set of matrices that does not depend on a
background metric.   This is easy to do, as the following 
example illustrates.

A theory depending on $d$, $N \times N$ matrices, $X_a$, $a=1,...,d$
that does not depend on a background metric 
is described by the action,
\f
S^d= \epsilon^{a_1...a_d}Tr\left [  X_{a_1}...X_{a_d}  \right ]
\ff
This vanishes trivially for even $d$, as a result of the
Jacobi identity.  This simple fact is
analogous to the fact that in the continuum
\f
S^d_{cont}= \int Tr   \left [ F  \wedge F  ... \wedge F \right ]
\ff
is a topological invariant,
as the Bianchi identity reduces to the Jacobi identity of the
matrices.
But for odd $d$ the action $S^d$ does not vanish.  Instead, one has
a kind of matrix analogue of Chern-Simons theory.
Interestingly, higher dimensional Chern-Simon theories
have local degrees of freedom\cite{mecs11,mhcs}, and the structure
of their constraints and equations of motion can be intricate.  

For odd $d=2n+1$ the
equations of motion are,
\f
{\delta S^{2n+1} \over \delta X_a}=
\epsilon^{a b_1...b_{d-1}} X_{b_1}...X_{b_{d-1}} =0
\ff
The solution spaces of these theories include the solution
spaces of the background dependent theories is which
$[X_a, X_b]=0 $ for all $a ,b$.  At the same time, the global symmetry
of the background dependent matrix models,
$X_a \rightarrow X^\prime_a = X_a + V_a I$, where $I$ is
the identity matrix and $V_a$'s are constants, is replaced
by a {\it gauge invariance}
\f
X_a \rightarrow X^\prime_a = X_a + V_a (X) I
\label{gauge}
\ff
where the $V_a (X)$ are now {\it functions} on the space
of matrices.   To see this note that,
\f
\delta S^{2n+1}= \epsilon^{a b_1...b_{2n}}
V_a Tr\left [  X_{b_1}...X_{b_{2n}}  \right ]=0
\ff

We can see these features as well from the canonical formalism.
We may introduce a continuous time by represnting explicitly the time
dimension.  The $2n+1$ component we represent as
time, so we write $X_{2n+1}=A_0$.  We then have
\f
S^{\prime 2n+1}= \int ds \epsilon^{b_1...b_{2n}}
V_a Tr\left [  ({\cal D}_0 X_{b_1}) X_{b_2}...X_{b_{2n}}  \right ]=
\ff
The canonical momenta are,
\f
\Pi^a = \epsilon^{ab_2...b_{2n}}
\left [  X_{b_2}...X_{b_{2n}}  \right ]
\ff
There is a gauge constraint,
\f
G= [X_a , \Pi^a ]
\ff
In addition, there are $2n$ constraints,
\f
D^a = Tr\Pi^a =0
\ff
that follow from the vanishing of $S^{2n}$.
These generate the $2n$, ``spatial" components
of the gauge symmetry (\ref{gauge}).

More structure
may be introduced by following the strategy of CDJ \cite{cdj}
and introducing lagrange multipliers into the action.
This will be discussed elsewhere.

Of course, this is not the only possible
approach to a background independent dynamics of matrices.
The new path integral formulations of spin network
evolution may be interpreted as a dynamics for matrices, if the
spin networks are taken to be not embedded in any background
manifold, as is advocated in \cite{fotini1}.  Non-embedded
spin networks are equivalent to a set of matrices, which are
constructed from their adjacency matrices\cite{fmpc}.  
Of course, the relevance of any of these models to $\cal M$ theory
remains to be shown.  

\section{Conclusions}

Put briefly, we have made some progress towards a covariant
formulation of membrane dynamics.  The crucial issues left so
far unsolved are,

\begin{itemize}

\item{}The choice of the physical inner product for the physical 
states, which is unlikely to be the same as the in the light-cone
gauge fixed theory.  This opens up the issue of the physical
interpretation of the quantum states of the membrane as well
as the consistency of a non-perturbative quantization of the
membrane in any dimensions.   

\item{}The possibility of a matrix representation of the covariant
theory rests on the implementation of a non-linear realization of
the non-area preserving diffeomorphisms of the membrane.  This
gauge symmetry, together with the hamiltonian constraint, is
necessary to balance the increase in the number of matrices from
$d-2$ to $d$ which moving from the light cone gauge to a covariant
formalism requires.  

\end{itemize}

Further work in this subject will also include the extension to
the supermembrane, which will involve also the study of special
dimensions such as $d=10+1$.  But the results found so far in this
general study tell us what those more specific studies will have
to accomplish if there is to be a Lorentz covariant formulation of
$\cal M$ theory arising from the dynamics of membranes.

\section*{ACKNOWLEDGEMENTS}

I am grateful first of all to Shyamoli Chaudhuri, Miao Li 
and Djordje  Minic for many discussions about the matrix models
and to Fotini Markopoulou for advice about the canonical formalism.
Discussions with Eli Hawkins, Murat Gunyadin, Renata Kallosh and 
Carlo Rovelli were also very helpful.  This work was supported by
NSF grant PHY-9514240 to The Pennsylvania State
University.


\begin{thebibliography}{99}


\bibitem{smem}E. Bergshoeff, Szegin and P. Townsend,
Ann.  Phys. 185 (1988) 330; B. de Wit, M. Luscher and H.
Nicolai, Nucl. Phys. B305[FS23] (1988) 545.

\bibitem{mtheory}E.  Witten, {\it String theory dynamics
in various dimensions} Nucl. Phys. B443 (1995) 85, hep-th/9503124.

\bibitem{DHN}B. DeWitt, J. Hoppe, H. Nicolai, Nuclear Physics B305
(1988) 545.

\bibitem{BFSS}T. Banks, W. Fischler, S. H. Shenker and L. Susskind,
hep-th/9610043, Phys. Rev. D55 (1997) 5112.

\bibitem{mym}R. Dijkgraaf, E. Verlinde and H. Verlinde,
hep-th/9703030.

\bibitem{ml}N. Ishibashi, H. Kawai, Y. Kitazawa and T. Tsuchiya, 
hep-th/9612115.

\bibitem{early}S. A. Godilidze, V. V. Sanadze, Yu. S. Surovtsev,
F. G. Tkebuchava, Tbilisi preprint E2-89-243 (1989).

\bibitem{zero}M. Porrati and A. Rozenberg, hep-th/9708119,
J. Hoppe hep-th/9709132, J. Frohlich and J. Hoppe, hep-th/9701119.

\bibitem{limitstrings}E. Bergshoeff, Szegin and P. Townsend,
Ann.  Phys. 185 (1988) 330.
U. Lindstrom, Phys. Lett. B 218 (1989) 315;

\bibitem{kodama}H. Kodama, Phys. Rev. D 42 (1990) 2548.

\bibitem{cstate}B. Bruegmann, R. Gambini and J. Pullin,
Phys. Rev. Lett.  68,  431 (1992); Gen. Rel. and Grav. 251 (1993);
R. Gambini and J. Pullin, 
{\it  Loops, knots, gauge theories and quantum gravity}
Cambridge University Press, 1996.

\bibitem{chopinme}L. Smolin and C. Soo,  
{\it The Chern-Simons invariant as the natural time variable for
classical and quantum gravity}  
Nucl. Phys. B 327 (1995) 205.

\bibitem{mejmp}L. Smolin, {\it Linking topological quantum
field theory and nonperturbative quantum gravity}
gr-qc/9505028,   J. Math. Phys. 36 (1995) 6417.

\bibitem{mecs11}L. Smolin, {\it Chern-Simons theory in 11 dimensions 
as a non-perturbative phase of M theory},
preprint,  revised, October 1997.

\bibitem{mhcs}M. Banados, M. Henneaux, C. Iannuzzo and C. M.
Viallet, {\it A note on the gauge symmetries of pure Chern-Simons
theory with p-form gauge fields} gr-qc/9703061.
\bibitem{BB}J. B. Barbour, Nature 249 (1974) 
328 (Erattum Nature 250 (1974) 606;
Nuovo Cimento 26B (1975) 16;
J. B. Barbour and B. Bertotti, Nuovo Cimento 38B (1977) 1;
Proc. Roy. Soc.   Lond. A 382 (1982)  295.

\bibitem{BB-me}L. Smolin {\it Space and Time in the Quantum
Universe}  in {\it Conceptual Problems of 
Quantum Gravity} ed. by A. Ashtekar and J. Stachel,
(Birkhauser,Boston,1991); J. B. Barbour and L. Smolin {\it Can quantum
mechanics be applied to the universe as a whole?} Yale University 
preprint, (1987).

\bibitem{me8}L. Smolin {\it Octonionic formulation of
$\cal M$ theory} preprint in preparation.

\bibitem{cdj}R. Capovillia, J. Dell and T. Jacobson,
Phys. Rev. Lett. 63 (1989) 2325.

\bibitem{lp1}T. Jacobson and L. Smolin, Nucl. Phys. B 299 (1988); 
C Rovelli L Smolin: Phys Rev Lett 61 (1988) 1155; Nucl 
Phys B133, 80 (1990).

\bibitem{volume1}C. Rovelli and L. Smolin
{\it Discreteness of area and volume in quantum gravity}
 Nuclear Physics B 442 (1995) 593.  Erratum: Nucl. Phys.
B 456 (1995) 734.

\bibitem{future}L. Smolin, {\it The future of spin networks}
 gr-qc/9702030.

\bibitem{rigorous}A. Ashtekar and C. J. Isham, 
 Class and Quant  Grav 9 (1992) 1069;
A Ashtekar J Lewandowski D Marlof J 
Mour\~{a}u T Thiemann:  ``Quantization of diffeomorphism
invariant theories of connections with local degrees of
freedom", gr-qc/9504018, JMP 36 (1995) 519;
A. Ashtekar and J. Lewandowski, "Quantum
Geometry I: area operator" gr-qc/9602046; 
J. Lewandowski, "Volume and quantization"
gr-qc/9602035.

\bibitem{duality}M. J. Duff, P. Howe, T. Inami, K.S. Stelle, {\it 
Superstrings in D=10 from supermembranes in D=11}, 
Phys. Lett. B191 (1987) 70; M. J. Duff, J. X. Lu {\it Duality 
rotations in membrane theory}, Nucl. Phys. B347 (1990) 394;
M. J. Duff, R. Minasian, James T. Liu {\it Eleven dimensional
origin or string/string duality: a one-loop test}, 
Nucl. Phys. B452 (1995) 261; C. Hull and P. K. Townsend,
{\it Unity of superstring dualities}, Nucl. nPhys. B438 (1995) 109,
hep-th/9410167; {\it Enhanced gauge symmetries in
superstring theories} Nucl. Phys. B451 (1995) 525, hep-th/9505073;
P. K. Townsend, {\it The eleven dimensional supermembrane
revisited} Phys. Lett. B350 (1995) 184, hep-th/9501068;
{\it String-membrane duality in seven dimensions}, Phys.
Lett. 354B (1995) 247, hep-th/9504095.

\bibitem{instability}L. Smolin, {\it  The classical
limit and the form of the hamiltonian constraint in non-pertubative
quantum gravity} CGPG preprint, gr-qc/9609034.

\bibitem{fmls}F. Markopoulou and L. Smolin, {\it Causal evolution of
spin networks} gr-qc/9702025.  CGPG 97/2-1;

\bibitem{fotini1}F. Markopoulou,  
{\it Dual formulation of spin network evolution}
preprint, March 1997.

\bibitem{fmpc}F. Markopoulou, in preparation.

\end{thebibliography}
\end{document}